\documentclass[10pt]{article}

\usepackage{epsfig}

\newlength{\dinwidth}
\newlength{\dinmargin}
\marginparpush
5pt \topmargin -42pt \headheight 12pt

\newcommand{\aver}[1]{\langle \, #1 \, \rangle \mathstrut}

\newcommand{\fr}[2]{\left(\frac{#1}{#2}\right)}

\begin{document}
\titlepage

\vspace*{4cm}

\begin{center}

{\Large \bf DM clumps as discrete sources of gamma-radiation}

\vspace*{1.5cm}
\textsc{K.M. Belotsky$^{a,b}$, A.M. Galper$^{a}$}

\vspace*{0.5cm}
$^a$ Moscow Engineering Physics Institute, \\ Kashirskoe
shosse 31, 115409,
Moscow, Russia \\[0.5ex]
$^b$ Center for Cosmoparticle Physics "Cosmion" of
Keldysh Institute of Applied Mathematics, \\
4 Miusskaya pl., 125047, Moscow, Russia \\[0.5ex]

\vspace*{0.5cm}

\end{center}

\begin{abstract}
Massive objects (clumps) of Cold Dark Matter (CDM) in Galaxy can appear due to its annihilation as discrete sources of gamma-radiation.
Some number of unidentified regular gamma-sources, observed by EGRET, can be accounted for by massive CDM clumps.
Future gamma-ray expreriment GLAST in combination with data of EGRET will enable to probe a wide range of models of 
clumped annihilating CDM.
\end{abstract}

\vspace*{1cm}

Cold Dark Matter (CDM) can be searched for indirectly by effects of its annihilation. 
%Annihilation of CDM, distributed in Galaxy, can give contribution into cosmic rays (CR), isotropic gamma radiation.
Some part of CDM distributed in Galaxy can be in form of small scale inhomogeneities (clumps).
The latters can play important role in searches of CDM.
Clumps are predicted to be the extent objects with density on a few orders of magnitude in averaged greater than
that of ambient CDM while being inhomogeneously distributed: it grows strongly from clump periphery to its center.
Due to high clump interior density, clumps can provide a strong enhancement of the annihilation signals, facilitating
their search in cosmic rays (CR), diffuse gamma radiation.
Furthermore, a sinlge clump, which annihilation flux is formed mainly inside small its central part, can be observed
as a regular discrete gamma-source.
Definite prediction of observable features of these sources requires a specification of a large set of parameters concerning 
both inhomogeneities formation and CDM particles properties. 
In this notice the only choice of parameters is taken to assess the possibility for EGRET to have registered such clumps
among discrete gamma-sources (see \cite{EGRET}) and the possibility for GLAST to probe them in future.
%
%\begin{figure}
%\special{center EGRET.JPG, \the\hsize 6cm 6cm 6cm}
%\vspace{7cm}
%\end{figure}
%
%\begin{figure}
%\special{center EGRET.jpg, \the\hsize 10cm 6cm 6cm}
%\vspace{10cm}
%\caption{Map of gamma-sources measured by EGRET \cite{EGRET}.} 
%\end{figure}

Density profile inside clumps is predicted to be in form of power law $\rho(r)\propto 1/r^{\beta}$
with cut off of density growth at some small radius of core $r_c$. We will adopt following \cite{Berez} 
\begin{equation}
\rho(r)=\left\{
     \begin{array}{l}
     \rho_c,\;\;\; r<R_c\\
     \rho_c\fr{R_c}{r}^{\beta},\;\;\; R_c<r<R\\
     0,\;\;\;  r>R,
     \end{array}
\right.
\label{rho}
\end{equation}
taking the radius of clump of mass $M$ to be $R=10^{18}\fr{M}{M_{\odot}}^{1/3}$ cm, $\beta=1.8$.
Further the ratio $M/M_{\odot}$ between the clump and Solar masses will be denoted as $\hat{M}$.
Predictions for $R$ and $\beta$ vary in relatevely small range around the quoted magnitudes.
Core radius, being defined by $x_c=R_c/R$, has a large scatter in its predictions.
The work \cite{Berez}, where tidal interactions are taken into account, gives largest estimation for it, $x_c\sim 0.05$.
It is evident, that clumps with more sharp density profiles (smaller $x_c$, greater $\beta$) at the fixed their mass 
would provide more intense and concentrated sources.
In our estimation $x_c=0.05$ is put.
We will use given density profile (\ref{rho}) for a wide range of clump mass, outspread well farther than that in \cite{Berez},
being based on universality of this law predicted in \cite{Gurevich}.

Let the CDM particle ($X$) be Majorana fermion with mass $m=100$ GeV 
(in case of Dirac fermion one needs to take into account that $\rho_{X}=\rho_{\bar{X}}=\frac{1}{2}\rho$, 
provided a charge symmety of $X$ and $\bar{X}$), 
having annihilation cross section (predominantly in s-wave)
\begin{equation}
\aver{\sigma v}=2\cdot 10^{-26}\,{\rm cm^3/s}.
\end{equation}
Given magnitude approximately provides a frozen out density of these particles required in cosmology.
Multiplicity of $\gamma$-quanta produced in one annihilation act above energy threshold of EGRET ($E_{\min}=100$ MeV)
is assumed to be $N_{\gamma}=20$, what roughly corresponds to typical high energy physics (HEP) processes 
with energy release $\sim 200$ GeV.

The flux of $\gamma$-quanta from a single clump at the distance to its center $l$ is given by
\begin{equation}
F_{\gamma}=\frac{N_{\gamma}\aver{\sigma v}}{4\pi l^2} \int \fr{\rho(r)}{m}^2dV.
\end{equation}
Integration over volume should be perfomed within solid angle corresponding to angle resolution of EGRET ($\delta=0.5^{\circ}$).
If given flux $F_{\gamma}>F_{\min}=3\cdot 10^{-8}$ cm$^{-2}$s$^{-1}$ then it could be recognized over background by EGRET.

Experiment GLAST, to be carried through since 2007, is planned to have $\delta=0.25^{\circ}$, 
sensitivity level for one year of operation $F_{\min}=10^{-10}$ cm$^{-2}$s$^{-1}$ at ehergy threshold 2 GeV.
For energy $E<2$ GeV sensitivity of GLAST $F_{\min}$ is expected to rise as $\propto E^{-1}$. What makes the use of threshold
$E_{\min}=2$ GeV be more favourable in case $N_{\gamma}(>E)$ behaves more smooth in this energy range. The latter is, as rule,
the case for HEP processes, and we will assume $N_{\gamma}(>2\,{\rm GeV})=5$.

\begin{figure}
\begin{center}
\centerline{\epsfxsize=8cm\epsfbox{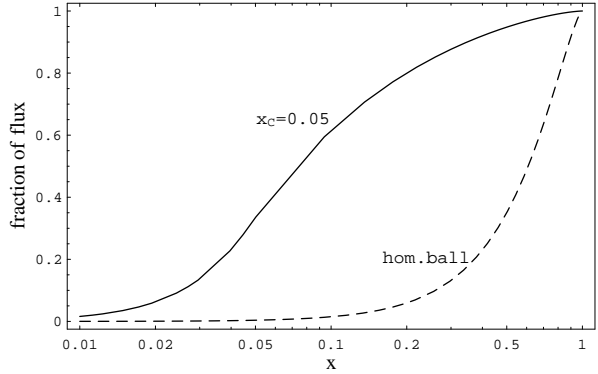}} 
\caption{Fraction of whole clump's flux which goes from central region of CPCS corresponding to its relative radius $x=r/R$
for density profile parameter $x_c=0.05$. For comparison the case of ball with uniform density
(homogenous ball) is shown too.} 
\label{flux}
\end{center}
\end{figure}

Because of finite (quite large) size of clump and its sharp density profile, 
at the small distances ($l<l_{\delta}=2R/\delta$) only central part of clump 
can be observed as a discrete source.
Indeed, the most of flux emitted by clump goes from a very small central part of clump projection on celestial sphere (CPCS).
For instance at $x_c=0.05$, $\beta=1.8$, one third of all flux from the clump goes from the region of CPCS enclosed
by radius $R_c$ around its center, which makes up 1/400 part of all projection area. 
Figure \ref{flux} shows which fraction of all clump's flux
goes from region of CPCS corresponding to its relative radius $x=r/R$.

At some smaller distances ($l<l_{\min}<l_{\delta}$), clump can not be seen as a pointlike source, gamma-flux going from
regions of angle size $\delta$, neighbouring analogous central region of CPCS, exceeds $F_{\min}$. 
Condition $F(3\delta)>6F(\delta)$, where arguments correspond to angle diameters with respect to clump center of regions being taken,
will be put as a criterion of non-pointlikeness of clump.

For estimation of probability to find a clump of mass $M$ within distance $l$ from the Earth, we assume such clumps
make up $\xi=0.002$ of total density of matter in our neighbourhood, being estimated as $\rho_{loc}\sim 0.3$ GeV/cm$^3$.
Given $\xi$ corresponds to the estimates for the clumps, survived to the present time \cite{Berez}.
Under our assumption the value
\begin{equation}
a=\frac{1}{^3\!\! \sqrt{n_{cl}}}=\fr{M}{\xi\rho_{loc}}^{1/3}\approx 40 \hat{M}^{1/3}{\rm pc}
\label{a}
\end{equation}
will give the mean distance between the clumps.
Within radius $l$ around the Earth there might be about
\begin{equation}
N_0= \frac{4}{3}\pi \fr{l}{a}^3 {\rm clumps}.
\label{N}
\end{equation}

\begin{figure}
\begin{center}
\centerline{\epsfxsize=11cm\epsfbox{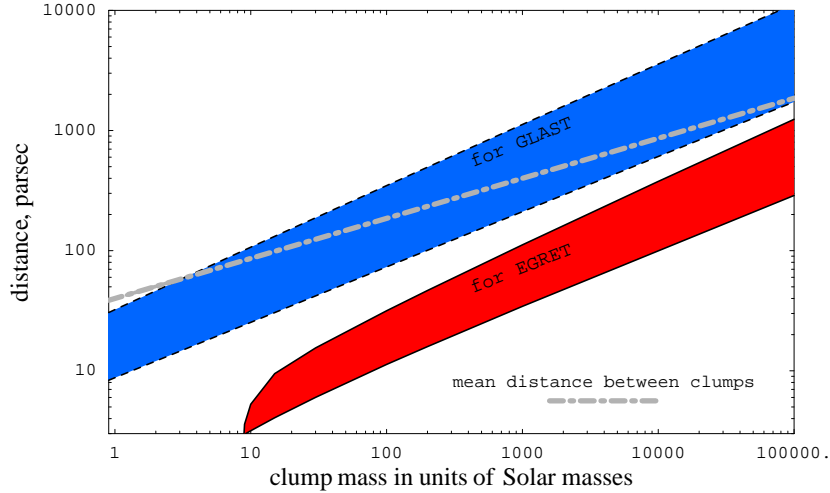}} 
\caption{Intervals of distances at which single clump can be observed as discrete gamma-source. 
The interval for EGRET (for GLAST) is enclosed between solid (dush) lines and is coloured by red (blue). 
Below lower limit of interval (coloured region), clump will look as a non-pointlike source. 
Mean distance between the clumps, as estimated in text, is shown too.}
\label{distance}
\end{center}
\end{figure}

On the figure \ref{distance} distances at which a single clump can be seen by EGRET and GLAST
as a pointlike $\gamma$-source are shown. At given $M$ there exist a maximal distance $l_{\max}$ below which
a clump becomes seen (upper limit of coloured region) and a minimal distance $l_{\min}$ below which
a clump is seen as non-pointlike (lower limt of coloured region).
Mean distance between the clumps Eq.(\ref{a}) are put on Fig.\ref{distance} too.
Note, that the distance $l_{\delta}$ at which a clump takes $0.5^{\circ}$ on the sky is about 2 times larger than $a$ 
(at any $M$). So, the most of curves present on Fig.\ref{distance} relate to gamma-emission of only central parts of clumps.

From figure \ref{distance} one can make a few notes, which relate to the chosen set of parameters.
EGRET is sensitive to only massive clumps, with $M>10 M_{\odot}$. GLAST will be able to observe
as $\gamma$-sources the clumps being 10 times farther than EGRET could observe, 
and respectively to observe 1000 times greater their amount (if EGRET did observe some of them). 
Intervals of distances at which clump 
can be registered by EGRET and by GLAST as discrete $\gamma$-source do not intersect.
So, if EGRET did observe clump(s) then it(they) should be registered as non-pointlike $\gamma$-source(s)
by GLAST.

Note, that $l_{\max}$ shown on Fig.\ref{distance} depends on parameter $x_c$ approximately as $\propto x_c^{-\beta+3/2}$.
For $x\ll 0.05$, what can be true for heavy clumps, $l_{\max}$ increases, threshold in $M$ for EGRET decreases,
regions for EGRET and GLAST can overlap.

EGRET detected 170 unidentified $\gamma$-sources \cite{EGRET}. 
The most of the sources is clear to be distributed anisotropically, 
concentrating to Galactic center and plane. Since, length scale of question,
given for clumps by Fig.\ref{distance}, is small relative to characteristic Galactic length scale, then 
$\gamma$-sources connected with CDM clumps can hardly exhibit such an anisotropy
(especially an anisotropy with respect to Galactic disk). Therefore, considered CDM clumps 
would be most likely to account for a possible (small) isotropic population of unidentified sources. 
The main, anisotropic part of $\gamma$-sources requires a separate explanation;
for instance, it can be connected with Wolf-Rayet stars \cite{WR} 
(also on possible origin of $\gamma$-sources see \cite{05}).
Roughly one supposes that the isotropic population numbers $\sim 10$ $\gamma$-sources.
%Neglecting by small fraction of CDM clumps which should look as non-pointlike,
For number of Eq.(\ref{N}) one requires $N_0(l_{\max})-N_0(l_{\min})\approx N_0(l_{\max})\sim 10$, from where
$l_{\max}\sim a$. From Fig.\ref{distance} one obtaines $M\sim a\, few\times 10^5 M_{\odot}$.

One reminds a degree of model dependence of given result. Many uncertain factors ($x_c$, $\beta$, $R$, $\xi$, distribution
of clumps in $M$, $m$, $\aver{\sigma v}$, $N_{\gamma}$ etc.) affects the result 
(some of which were taken here to be more optomistic) furthemore many of them are purely theoretic.

Finally note, that annihilating CDM concentrated around black holes of star masses 
can be considered as sources of $\gamma$-radiation. A degree of contraction of CDM by black hole gravity
can be estimated with relationship $n/v=const$ between number density and velocity of particles inside potential field \cite{nv}
(but this gives averaged $n$ over spheres around centre of gravity and lead to underestimation of annihilation rate if
density distribution over each sphere is inhomogeneous).
However, integral $\gamma$-flux turns out to be so small
that it seems incredible to make it observable at the expected distances.

\end{document}